\documentclass[runningheads]{llncs}
\usepackage[T1]{fontenc}
\usepackage{graphicx,verbatim}
\usepackage{tabularx}  

\usepackage{cite}
\usepackage{amssymb}
\usepackage{makecell}
\usepackage{enumerate}
\usepackage{threeparttable}
\usepackage{xcolor}
\usepackage{color}
\usepackage[colorlinks=true, linkcolor=red, urlcolor=blue]{hyperref}
\usepackage{marvosym}
\begin{document}
\title{Accurate and Efficient Fetal Birth Weight Estimation from 3D Ultrasound}
%

\author{Jian Wang\inst{1,2,3}\thanks{Jian Wang, Qiongying Ni and Hongkui Yu contribute equally to this work.\\
Corresponding email:
\email{xinyang@szu.edu.cn} and
\email{lujing810@foxmail.com}}\and Qiongying Ni\inst{1,3\star}\and Hongkui Yu\inst{5\star}\and  Ruixuan Yao\inst{6}\and Jinqiao Ying\inst{7}\and  Bin Zhang\inst{5}\and  Xingyi Yang\inst{5}\and Jin Peng\inst{1,3}\and  Jiongquan Chen\inst{1,3}\and Junxuan Yu\inst{1,3}\and Wenlong Shi\inst{8}\and Chaoyu Chen\inst{1,3}\and Zhongnuo Yan\inst{1,3}\and Mingyuan Luo\inst{1,3}\and  Gaocheng Cai\inst{8}\and Dong Ni\inst{1,2,3}\and Jing Lu\inst{4}\textsuperscript{(\Letter)}\and Xin Yang\inst{1,3}\textsuperscript{(\Letter)}
}  

\authorrunning{Jian Wang, Qiongying Ni and Hongkui Yu et al.}
\institute{
$^1$National-Regional Key Technology Engineering Laboratory for Medical Ultrasound, School of Biomedical Engineering, Shenzhen University, Shenzhen, China\\
$^2$College of Computer Science and Software Engineering, Shenzhen University, Shenzhen, China\\
$^3$Medical Ultrasound lmage Computing (MUSIC) Lab, School of Biomedical Engineering, Medical School, Shenzhen University, China\\
$^4$The First Afﬁliated Hospital of Xiamen University, School of Medicine, Xiamen University, Xiamen, China.\\
$^5$Shenzhen Baoan Women's and Children's Hospital, Shenzhen, China.\\
$^6$University of California, Berkeley, Berkeley, CA, USA\\
$^7$Yongkang Maternal and Child Health Hospital, Yongkang, China\\ 
$^8$Shenzhen RayShape Medical Technology Co., Ltd, Shenzhen, China\\
}
\maketitle               
\begin{abstract}
Accurate fetal birth weight (FBW) estimation is essential for optimizing delivery decisions and reducing perinatal mortality. However, clinical methods for FBW estimation are inefficient, operator-dependent, and challenging to apply in cases of complex fetal anatomy. Existing deep learning methods are based on 2D standard ultrasound (US) images or videos that lack spatial information, limiting their prediction accuracy. In this study, we propose the first method for directly estimating FBW from 3D fetal US volumes. Our approach integrates a multi-scale feature fusion network (MFFN) and a synthetic sample-based learning framework (SSLF). The MFFN effectively extracts and fuses multi-scale features under sparse supervision by incorporating channel attention, spatial attention, and a ranking-based loss function. SSLF generates synthetic samples by simply combining fetal head and abdomen data from different fetuses, utilizing semi-supervised learning to improve prediction performance. Experimental results demonstrate that our method achieves superior performance, with a mean absolute error of $166.4\pm155.9$ $g$ and a mean absolute percentage error of $5.1\pm4.6$\%, outperforming existing methods and approaching the accuracy of a senior doctor. Code is available at: \url{https://github.com/Qioy-i/EFW}.


\end{abstract}
\section{Introduction}
\begin{figure}
\includegraphics[width=\textwidth]{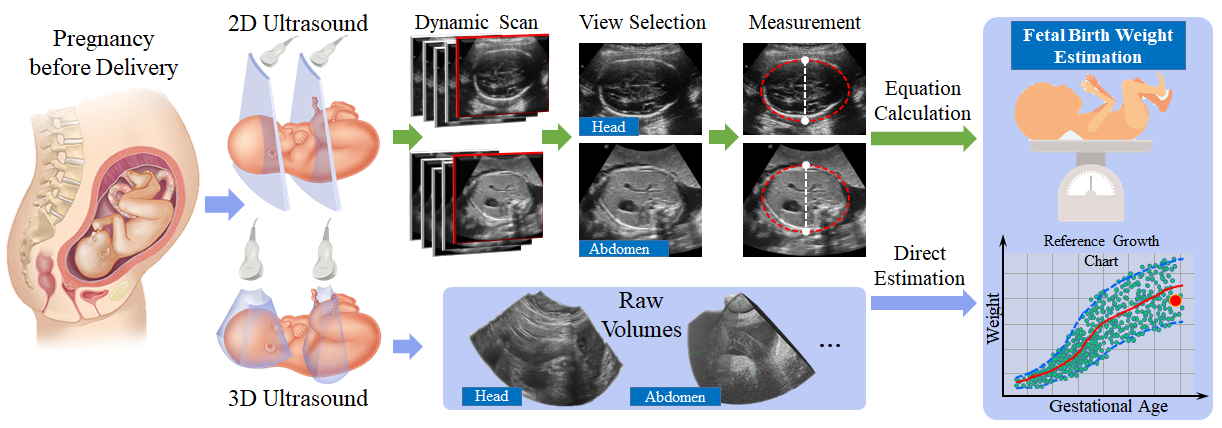}
\caption{Workflow for conventional 2D US-based and our 3D US-based FBW estimation.} \label{fig1}
\end{figure}

Accurate estimation of fetal birth weight (FBW) prior to delivery is crucial for making optimal delivery decisions (i.e., vaginal birth or cesarean section) and reducing perinatal mortality \cite{milner2018accuracy}. As shown in Fig.~\ref{fig1}, the standard clinical method for FBW assessment involves a cumbersome procedure: obtaining 2D standard ultrasound (US) planes, measuring biometric parameters, and applying an empirical formula to estimate it \cite{hadlock1985estimation}.
However, each steps poses significant challenges.
First, acquiring standard US planes requires radiologists to deeply understand fetal anatomy and accurately locate the 2D plane in 3D space, which is technically demanding and time-consuming \cite{chen2015automatic}.
Second, both standard plane acquisition and biometric measurements depend heavily on the radiologist’s experience, leading to intra- and inter-observer variability, which affects the accuracy and reproducibility \cite{li2025rule}. Finally, the empirical formulas currently in use are based on regression analysis of a limited set of biometric parameters, failing to account for diverse and complex anatomical variations among different fetuses \cite{ma2023evaluating}. \par

To address these clinical challenges, several artificial intelligence-based approaches have been proposed, broadly categorized into three types: 
(1) Some studies use manually measured parameters and maternal information, applying machine learning methods to predict FBW \cite{lu2020prediction, tao2021fetal, camargo2023multimodal}. 
(2) Other studies leverage deep learning to predict parameters from standard US planes or videos, followed by applying empirical formulas to estimate FBW \cite{bano2021autofb, plotka2021fetalnet, plotka2022deep}. 
(3) A few studies employ deep learning models to directly predict FBW from standard US videos combined with clinical information \cite{plotka2022babynet, plotka2023babynet++, plotka2023deep}. 
However, to date, no approaches have been developed to comprehensively address all three key challenges. 
Furthermore, due to the lack of spatial information in 2D US imaging, methods based on it struggle to accurately capture fetal anatomical structures, limiting prediction accuracy. 
In contrast, as shown in Fig.~\ref{fig1}, 3D fetal US imaging provides more comprehensive anatomical information and reduces reliance on the operator’s technique, making it a promising alternative for estimating FBW \cite{kang2021predicting, metwally2022three, hamza2025accuracy}. 
However, its relatively lower resolution and higher noise levels pose challenges for manual biometric measurements and empirical formula-based estimation. Moreover, while 3D US has not yet achieved widespread clinical adoption due to factors like cost and accessibility, it is increasingly available in advanced obstetric settings. This trend highlights the growing need for automated methods that can fully leverage 3D US data and bypass the limitations of handcrafted features and empirical modeling. \par

In this study, we propose the first method for directly estimating FBW from 3D fetal US volumes. The key challenges lie in extracting informative features from large, anatomically complex 3D data under sparse supervision, and addressing data scarcity given the wide prediction space. To tackle these, we introduce a multi-scale feature fusion network (MFFN) that integrates head and abdomen volumes using channel attention and Mamba-based spatial attention. We further design a ranking-based loss based on inter-fetal FBW relationships to improve generalization. To alleviate limited sample size, we propose a synthetic sample-based learning framework (SSLF), which generates new fetal samples by mixing head and abdomen volumes from different fetuses and learns from them via semi-supervised training.
The experimental results demonstrate that our method achieves superior performance, exceeding existing methods and approaching the accuracy of a senior doctor. \par

\section{Methodlogy}
\begin{figure}
\includegraphics[width=\textwidth]{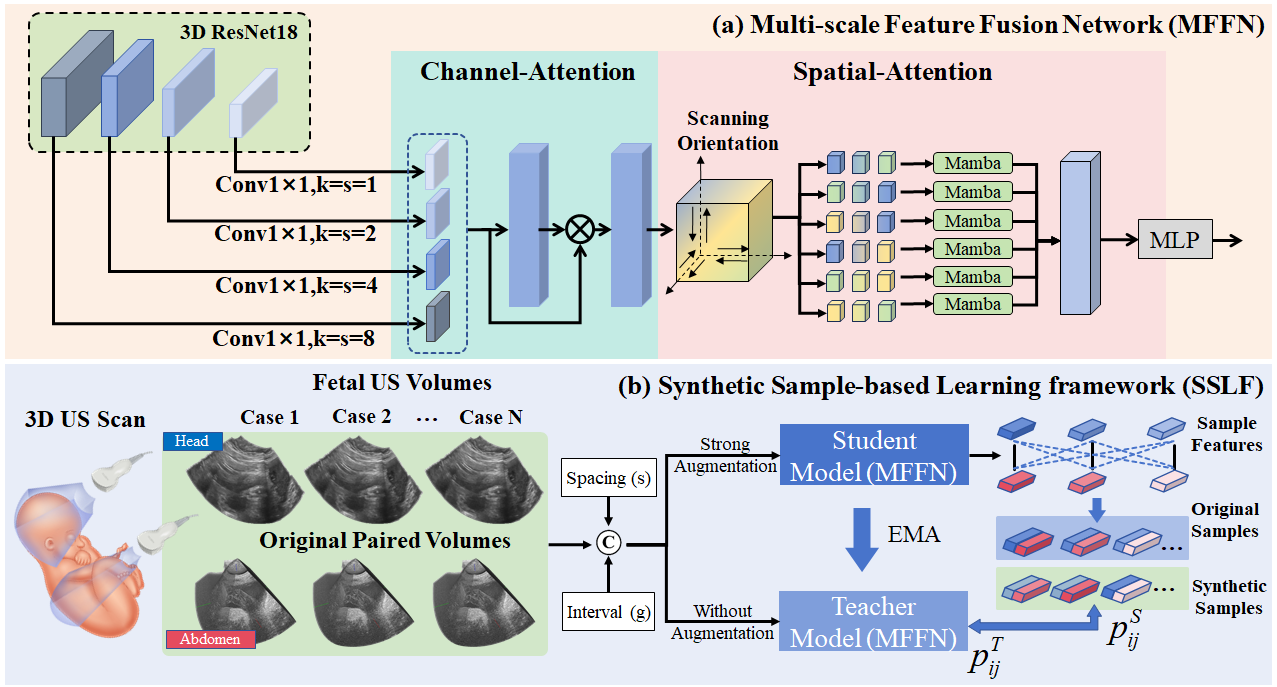}
\caption{Overview of the proposed FBW estimation method.} 
\label{fig_framework}
\end{figure}

As illustrated in Fig.~\ref{fig_framework}, our approach consists of a multi-scale feature fusion network (MFFN) and a synthetic sample-based learning framework (SSLF). In MFFN, we integrate channel attention and spatial attention to extract and fuse multi-scale features. Additionally, we design a ranking-based loss function to model the FBW relationships between fetuses. In SSLF, we design a simple yet effective sample synthesis method to expand the dataset and then apply semi-supervised learning to enhance FBW estimation.

\subsection{Multi-scale Feature Fusion Network}
Due to the significant variations in the spatial scales of fetal anatomical structures, features extracted from a single scale are inadequate for accurate FBW prediction. Thus, we design a multi-scale feature fusion network (Fig.\ref{fig_framework}(a)).

\textbf{Problem definition}: Let $\left \{ \left (  x_{i}^{h},x_{i}^{a},s_{i}^{h},s_{i}^{a},g_{i},y_{i}\right )|i\in 1,2,...N  \right \} $ denote a mini-batch randomly sampled from the 3D fetal US dataset. Here, $x_{i}^{h}, x_{i}^{a} \in \mathbb{R}^{1 \times D \times H \times W}$ represent the 3D US volumes of the fetal head and abdomen, where $D$, $H$, and $W$ are the depth, height, and width of the volumes, respectively. $s_{i}^{h}, s_{i}^{a} \in \mathbb{R}^{3}$ are the voxel spacing in the depth, height, and width directions for the head and abdomen. $g_{i} \in \left \{ 0, 1, 2, 3 \right \} $ is the time interval (in days) from the 3D US scan to delivery. $y_{i} \in [0, 1]$ denotes the normalized true FBW, and $N$ is the batch size. During clinical inference, $g_{i}$ is provided by the doctor as the time interval from the 3D US scan to the estimated delivery. To incorporate spacing and time interval information, we expand the dimensions of $s_{i}^{h}$ and $g_{i}$ and concatenate them along the channel dimension with $x_{i}^{h}$, resulting in a five-channel head volume $X_{i}^{h} \in \mathbb{R}^{5 \times D \times H \times W}$. Similarly, we obtain abdominal volume data $X_{i}^{a}$. The objective of this study is to predict FBW $y_{i}$ given the input volumes $(X_{i}^{h}, X_{i}^{a})$.

\textbf{Feature Fusion}: We designed a shared network for both head and abdomen data, so their forward computation processes are identical. For simplicity, we describe the computation process using head data as an example. 3DResNet18 \cite{tran2018closer} is adopted as the backbone for feature extraction. 
The features from four intermediate layers, denoted as $ Z_{i}^{h^1}, Z_{i}^{h^2}, Z_{i}^{h^3}, Z_{i}^{h^4} $, are extracted. These features are spatially downsampled by factors of 4, 8, 16, and 32, with corresponding channel dimensions of 64, 128, 256, and 512. To enable multi-scale feature fusion at a uniform spatial resolution, we first apply $1 \times 1 \times 1$ convolutions to project all feature maps to a consistent channel dimension of 128. Then, four sets of convolutions are applied to downsample the features to the same spatial resolution, with kernel sizes and strides set to 8, 4, 2, and 1, respectively. This process aligns features across all four scales both spatially and channel-wise. Finally, we concatenate the features along the channel dimension, resulting in the fused representation $Z_{i}^{h} \in \mathbb{R}^{512 \times \frac{D}{32} \times \frac{H}{32} \times \frac{W}{32}}$. \par

\textbf{Channel Attention}: High-level features typically capture global semantics but with lower resolution, while low-level features focus on fine-grained details at higher resolution. Given their different roles, it is necessary to weigh them appropriately. To achieve this, we apply channel attention to the fused features $Z_{i}^{h}$. We use a bottleneck MLP with the same input and output dimensions, where the activation function is ReLU and the dimensionality reduction factor is set to 16, denoted as $MLP(\cdot)$. The channel attention mechanism is formulated as:
\begin{equation}
Z_{i}^{h} =  Z_{i}^{h} \cdot \sigma (MLP(GAP(Z_{i}^{h}))),
\end{equation}
where $\sigma(\cdot)$ denotes the sigmoid activation function and $GAP(\cdot)$ represents global average pooling over the spatial dimensions. This attention mechanism assigns independent weights to channels at different scales, improving FBW prediction. \par

\textbf{Spatial Attention}: Since not all regions in the volume correspond to fetal anatomy, and different structures contribute differently to FBW prediction, it is essential for the model to focus on relevant fetal regions. Inspired by the Mamba architecture \cite{gu2023mamba}, we design a spatial attention mechanism that employs six Mamba models, each corresponding to a unique scanning orientation: (1) left-to-right, (2) right-to-left, (3) top-to-bottom, (4) bottom-to-top, (5) front-to-back, and (6) back-to-front. Specifically, six spatial sequences $Z_{i}^{h_1}, Z_{i}^{h_2}, Z_{i}^{h_3}, Z_{i}^{h_4}, Z_{i}^{h_5},$ $ Z_{i}^{h_6}\in \mathbb{R}^{512 \times L}$ are generated by applying different flattening methods to the input features, where $L = \frac{D}{32} \times \frac{H}{32} \times \frac{W}{32}$. The sequences are processed by their corresponding Mamba models, and their outputs are fused using an averaging operation, yielding the final feature map $Z_{i}^{h} \in R^{512 \times L}$. This fusion captures spatial information from multiple perspectives, improving FBW prediction.

\textbf{Loss Function}: After the above three steps, we obtain the head and abdomen features $Z_i^h$ and $Z_i^a$. These features are concatenated along the channel dimension and subjected to global average pooling over the spatial dimensions, resulting in the fused head-abdomen representation $Z_{i,i}^{ha} \in \mathbb{R}^{1024}$. A single fully connected layer followed by a sigmoid activation function is applied to predict the FBW $p_i$. The regression loss is defined as:
\begin{equation}
Loss_{reg} = \frac{1}{N} \sum_{i=1}^{N} MSE(p_{i},y_{i}),
\end{equation}
where $MSE(\cdot)$  denotes the mean squared error between the predicted FBW $p_i$ and the ground truth $y_i$. To further enhance the model's ability to capture relative weight differences among fetuses, we introduce a ranking-based loss function. This loss encourages the predicted order of FBWs to align with the ground truth order. The ranking loss is formulated as:
\begin{equation}
Loss_{rank} = \frac{1}{N^2} \sum_{i=1}^{N}\sum_{j=1}^{N} max[0,-(p_{i}-p_{j})]\cdot \tau (y_{i}-y_{j}),
\end{equation}
where $\tau (\cdot)$ equal to 1 if the argument is greater than 0 and 0 otherwise.

\subsection{Synthetic Sample-based Learning Framework}
In this study, a key challenge we face is the conflict between limited sample size and nearly infinite possible predictions, making conventional supervised learning models prone to severe overfitting. To address this issue, we propose an synthetic sample-based learning framework, as shown in Fig. \ref{fig_framework}(b).

\textbf{Sample synthesis}: Here, the sample refers to the fetus rather than the head or abdominal volume. We hypothesize that the model predicts FBW by estimating fetal size from both head and abdominal volumes. Due to the anatomical independence between head and abdominal regions, there is little or no direct correlation between their volumes within the same fetus. Based on this observation, we propose a simple method to synthesize new samples from different fetuses. Specifically, we create synthetic samples by combining the head volume of one fetus with the abdominal volume of another. Let $\left \{ \left (  X_{i}^{h},X_{j}^{a}\right )|i,j\in 1,2,...N , i\neq j\right \}$ denote the set of synthetic samples within a mini-batch, where $N$ is the batch size. For each mini-batch, this method can generate $N(N-1)$ synthetic samples, greatly expanding the dataset without any complicated operations. \par

\textbf{Semi-supervised learning}: Although the previous steps synthesize a large number of new samples to expand the dataset, they lack real FBW labels. To address this, we leverage semi-supervised learning by treating synthetic samples as unlabeled data. Inspired by the MeanTeacher framework \cite{tarvainen2017mean}, we design a semi-supervised learning setup consisting of two models: the teacher model and the student model, both of which share the same architecture (i.e., MFFN). During training, the student model's parameters are updated via backpropagation, while the teacher model’s parameters are updated using an exponential moving average (EMA) of the student model’s parameters, as follows:
\begin{equation}
\theta _{T}^{t} = m\theta _{T}^{t-1} + (1-m)\theta _{S}^{t} ,
\end{equation}
where $\theta_T$ and $\theta_S$ denote the parameters of the teacher and student models, respectively. $t$ denotes the iteration step, and $m$ is the momentum coefficient that controls the update rate of the teacher model.

In each training batch, there are $N$ head and $N$ abdomen data samples, resulting in $N^2$ possible combinations. Of these, $N$ combinations correspond to real fetal samples (with labels), while the remaining $N(N-1)$ combinations correspond to synthetic fetal samples (without labels). For these combinations, let $\left \{p_{ij}^{T} | i, j \in 1, 2, ..., N\right \}$ and $\left \{p_{ij}^{S} | i, j \in 1, 2, ..., N\right \}$ represent the predictions from the teacher and student models, respectively. We treat the teacher Model’s predictions as pseudo-labels to supervise the student Model. The loss function for the semi-supervised learning phase is defined as:
\begin{equation}
Loss_{semi}=\frac{1}{N(N-1)}\sum_{i=1}^{N}\sum_{j=1}^{N}MSE(p_{ij}^{S},p_{ij}^{T}), i\ne j. 
\end{equation}
Finally, the total loss function is the weighted sum of three components:
\begin{equation}
Loss_{total} = Loss_{reg} + \alpha Loss_{rank} + \beta Loss_{semi},
\end{equation}
where $\alpha$ and $\beta$ are the loss weighting coefficients. With this approach, our model can effectively learn from synthetic samples generated by combining fetal head data with abdomen data from different fetuses.

\section{Experiments and Results}

\subsection{Datasets and Implementations}

With ethics committee approval, we collected a dataset of 491 pregnancies by two senior doctors, each with 17 years of experience. Each case includes one head, abdomen, and femur ultrasound scan acquired within 72 hours before delivery, using a GE HealthCare E8 system with an RAB6-D probe. Prior to delivery, a senior doctor estimated fetal birth weight (FBW) using the Hadlock formula based on manual biometric measurements. Post-delivery, the true FBW was recorded as ground truth. The mean true FBW was $3229.3 \pm 467.8$ g, ranging from 1000 g to 4610 g. Voxel spacing was extracted from metadata. The dataset was randomly split into training, validation, and test sets at a 7:1:2 ratio, and all volumes were resized to $160 \times 128 \times 96$ while preserving aspect ratio.

We trained the model for 200 epochs with a batch size of 16 using the Adam optimizer. The learning rate started at $1\mathrm{e}^{-4}$, warmed up for 5 epochs, and then decayed via cosine annealing. Data augmentation included rotation, flipping, scaling, and contrast and brightness adjustments to improve generalizability. Hyperparameters were set as follows: $\alpha=0.001$, $\beta$ increased linearly from 0 to 0.2, and momentum $m$ from 0.99 to 0.9999. These values were selected via grid search on the validation set. All experiments ran on an RTX 4090 GPU server.

\subsection{Results}
To evaluate our method, we compared it with several existing approaches, including Hadlock~\cite{hadlock1985estimation}, INTERGROWTH-21\textsuperscript{st}\cite{stirnemann2017international}, BabyNet\cite{plotka2022babynet}, BabyNet++\cite{plotka2023babynet++}, 3DCNN-1\cite{wang20233dcnn}, 3DCNN-2~\cite{3DCNN-2}, and DSCNN~\cite{jafrasteh2023deep}.
Hadlock and INTERGROWTH-21st are widely used clinical methods based on manual biometric measurements and empirical formulas; we used the pre-delivery estimate by a senior doctor as the Hadlock reference.
BabyNet and BabyNet++ predict FBW from 2D ultrasound videos, with the latter also incorporating biometric and maternal data; we adopted the reported results from their original studies.
3DCNN-1, 3DCNN-2, and DSCNN are recent 3D regression models for other tasks, which we re-implemented following their published protocols.
All methods were evaluated using mean absolute error (MAE), root mean square error (RMSE), and mean absolute percentage error (MAPE).

Table~\ref{tab1} summarizes the comparison results. Our method achieves strong performance, with a MAE of $166.4\pm154.9$ g, RMSE of $227.3\pm227.3$ g, and MAPE of $5.1\pm4.6$\%. Notably, it is highly competitive with Hadlock—widely used in clinical practice—and INTERGROWTH-21\textsuperscript{st}, a global fetal growth standard. The closest method, BabyNet++, leverages standard US videos, manual biometric measurements, and maternal information. In contrast, our approach relies solely on 3D US volumes, requiring minimal scanning expertise and no manual input.
We also evaluated the two key components of our method. MFFN alone achieved a MAE of $172.0\pm164.6$ g, RMSE of $238.1\pm237.1$ g, and MAPE of $5.3\pm4.9$\%, demonstrating its ability to extract features effectively from 3D volumes under sparse supervision. When combined with SSLF, multiple 3D regression models showed improved performance, indicating its adaptability and potential for broader application.

\begin{table}[h!]
\centering  
\begin{threeparttable}
\caption{Comparison results of different methods.}\label{tab1}
\begin{tabular}{l @{\hspace{1cm}} c @{\hspace{1cm}} c @{\hspace{1cm}} c }
\hline
\textbf{Method} &  \textbf{MAE{(g)}}$\downarrow$ & \textbf{RMSE{(g)}}$\downarrow$ & \textbf{MAPE{(\%)}}$\downarrow$\\
\hline
Hadlock~\cite{hadlock1985estimation} & 160.2 ± 118.5 & 199.3 ± 197.9 & 5.0 ± 3.7\\
INTERGROWTH-21\textsuperscript{st}~\cite{stirnemann2017international} & 182.9 ± 137.0 & 228.5 ± 220.9 & 5.7 ± 4.2\\
BabyNet~\cite{plotka2022babynet}* & 254.0 ± 230.0 & 341.0 ± 215.0 & 7.5 ± 6.6\\
BabyNet++~\cite{plotka2023babynet++}* & 179.0 ± 19.0  & 203.0 & 5.1 ± 0.6\\
\hline
3DCNN-1~\cite{wang20233dcnn} & 268.7 ± 245.2 & 363.8 ± 358.7 & 8.2 ± 7.1\\
3DCNN-2~\cite{3DCNN-2} & 253.1 ± 222.1 & 336.8 ± 336.2  & 7.9 ± 6.9\\
DSCNN~\cite{jafrasteh2023deep} & 264.7 ± 239.1  &  356.7 ± 355.0 & 8.3 ± 7.8\\
\bfseries $MFFN_{(Ours)}$ & 172.0 ± 164.6 & 238.1 ± 237.1 & 5.3 ± 4.9\\
\hline
3DCNN-1 + SSLF & 264.7 ± 227.8 & 349.3 ± 348.9 & 8.2 ± 7.1 \\
3DCNN-2 + SSLF & 248.3 ± 212.2 & 326.6 ± 325.1  & 8.1 ± 6.7\\
DSCNN + SSLF &  220.0 ± 218.3 &  310.0 ± 309.1 & 6.8 ± 6.5\\
\bfseries $MFFN + SSLF_{(Ours)}$&  \textcolor{blue}{166.4 ± 154.9} & \textcolor{blue}{227.3 ± 227.3} & \textcolor{blue}{5.1 ± 4.6} \\
\hline
\end{tabular}
\begin{tablenotes}
\item*Results reported from original publication; model not retrained on our dataset due to input modality mismatch.
\end{tablenotes}
\end{threeparttable}
\end{table}

\begin{table}[h!]
\centering  
\caption{Experimental results of ablation study with different configurations of key components of our model. 3DR18: 3DResNet18, WS: Weight-sharing, FF: Feature-fusion, CA: Channel attention, SA: Spatial attention, RL: Ranking-based loss}\label{tab2}
\resizebox{\textwidth}{!}{
\begin{tabular}{c c c c c c c|c c|c c}
\hline
\multicolumn{7}{c}{\textbf{Model Architecture}} & \multicolumn{2}{|c|}{\textbf{Input}} &  \multicolumn{2}{c}{\textbf{Results}} \\
\hline
\makecell{3DR18} & \makecell{WS} & FF & \makecell{CA} & \makecell{SA} & \makecell{RL} & SSLF & Head & Abdomen & MAE{(g)} & MAPE{(\%)}\\
\hline
\checkmark &  &  &  &  & &  & \checkmark &  & 283.1 ± 229.0 & 9.0 ± 7.5\\
\checkmark &  &  &  &  & &  &  & \checkmark & 252.8 ± 191.8 & 8.0 ± 6.4\\
\checkmark &  &  &  &  & &  & \checkmark & \checkmark & 222.4 ± 193.2 & 6.7 ± 5.1\\
\checkmark & \checkmark &  &  & &  &  & \checkmark & \checkmark & 200.6 ± 178.1 & 6.1 ± 5.1\\
\checkmark & \checkmark & \checkmark &  &  &  &  & \checkmark & \checkmark & 189.7 ± 173.9 & 5.9 ± 5.4\\
\checkmark & \checkmark & \checkmark & \checkmark &  &  &  & \checkmark & \checkmark & 191.0 ± 166.1  & 5.9 ± 5.0\\
\checkmark & \checkmark & \checkmark & \checkmark & \checkmark &  &  & \checkmark & \checkmark & 177.2 ± 175.2 & 5.9 ± 5.0\\
\checkmark & \checkmark & \checkmark & \checkmark & \checkmark & \checkmark &  & \checkmark & \checkmark & 172.0 ± 164.6 & 5.3 ± 4.9\\
\bfseries\checkmark & \checkmark & \checkmark & \checkmark & \checkmark & \checkmark & \checkmark & \checkmark & \checkmark & 166.4 ± 154.9 & 5.1 ± 4.6 \\
\hline
\end{tabular}
}
\end{table}

To evaluate the effectiveness of each component in our method, we conducted a series of ablation experiments. Specifically, we assessed the impact of single-site and multi-site volumes, weight-sharing in multi-site volume fusion, feature fusion, channel attention, spatial attention, ranking-based loss function, and SSLF. The results, presented in Table \ref{tab2}, demonstrate that weight-sharing in multi-site volume fusion is crucial for accurate FBW estimation. Additionally, simple multi-scale feature fusion enhances accuracy, confirming its necessity. While channel attention increases the average MAE by 1.3$g$, it reduces the standard deviation by 7.8$g$, making it beneficial overall. Spatial attention further improves performance, decreasing the average MAE by 13.8$g$. The ranking-based loss function aligns the FBW relationships between fetuses with the groundtruth, effectively boosting prediction accuracy. Finally, SSLF significantly expands the dataset and improves performance through a semi-supervised learning framework. 

\section{Conclusion}
In this study, we present a novel method for predicting FBW directly from 3D fetal US volumes. To address challenges in feature extraction and sample size limitations, our method designs a multi-scale feature fusion network and a synthetic sample-based learning framework. Our results outperform existing methods and close to clinical accuracy. Rather than replacing 2D workflows, our method offers a complementary tool for enhanced decision-making. Future work will extend this framework to jointly estimate biometric parameters and FBW, improving automation and interpretability.

\begin{credits}
\subsubsection{\ackname} This work was supported by the grant from National Natural Science Foundation of China (82201851, 12326619, 62171290); Guangxi Province Science Program (2024AB17023); Yunnan Major Science and Technology Special Project Program (No.202402AA310052); Yunnan Key Research and Development Program (202503AP140014); Science and Technology Planning Project of Guangdong Province (2023A0505020002);Frontier Technology Development Program of Jiangsu Province (No. BF2024078).

\subsubsection{\discintname}
The authors have no competing interests to declare that are relevant to the content of this article.
\end{credits}

\bibliographystyle{splncs04}
\bibliography{mybibliography}







\end{document}